\documentclass{jpp}
\usepackage{graphicx}
\usepackage{natbib}
\usepackage[T1]{fontenc}
\usepackage{bm}
\usepackage{color}
\usepackage{graphicx}
\usepackage{amsmath}
\usepackage{amssymb}
\usepackage{amsfonts}
\usepackage{courier}
\usepackage{txfonts}
\usepackage{appendix}
\usepackage{subfigure}
\newcommand{\ud}{\mathrm{d}}

\ifCUPmtlplainloaded \else
  \checkfont{eurm10}
  \iffontfound
    \IfFileExists{upmath.sty}
      {\typeout{^^JFound AMS Euler Roman fonts on the system,
                   using the 'upmath' package.^^J}%
       \usepackage{upmath}}
      {\typeout{^^JFound AMS Euler Roman fonts on the system, but you
                   dont seem to have the}%
       \typeout{'upmath' package installed. JPP.cls can take advantage
                 of these fonts, if you use 'upmath' package.^^J}%
      }
  \else
  \fi
\fi

\ifCUPmtlplainloaded \else
  \checkfont{msam10}
  \iffontfound
    \IfFileExists{amssymb.sty}
      {\typeout{^^JFound AMS Symbol fonts on the system, using the
                'amssymb' package.^^J}%
       \usepackage{amssymb}%

      }{}
  \fi
\fi

\ifCUPmtlplainloaded \else
  \IfFileExists{amsbsy.sty}
    {\typeout{^^JFound the 'amsbsy' package on the system, using it.^^J}%
     \usepackage{amsbsy}}
    {\providecommand\boldsymbol[1]{\mbox{\boldmath $##1$}}}
\fi

\newcommand{\unit}[1]{\ensuremath{\bm{\widehat{\mathrm{#1}}}}}

\newcommand{\Zeff}{Z_{\rm eff}}

\usepackage[T1]{fontenc}
\usepackage{bm}
\usepackage{color}
\usepackage{graphicx}

\newcommand{\sub}[1]{\ensuremath{_{\text{#1}}}}
\newcommand{\rd}{\ensuremath{\mathrm{d}}}

\usepackage{ae}
\usepackage{amsmath}
\usepackage{amssymb}
\usepackage{amsfonts}
\usepackage{mathrsfs}
\usepackage{units}
\usepackage{natbib}

\newcommand{\be}{\begin{displaymath}}
\newcommand{\ee}{\end{displaymath}}
\newcommand{\bn}{\begin{equation}}
\newcommand{\en}{\end{equation}}

\usepackage{enumerate}

\newcommand{\nofrac}[2]{#1/#2}

\newcommand{\Eceff}{\ensuremath{E_{\rm c}^{\rm eff}}}

\newcommand{\Ec}{E_{\rm c}}

\newcommand{\netot}{n_{\rm e}^{\rm tot}}
\newcommand{\nef}{n_{\rm e}}

\newcommand{\nuSBar}{\bar{\nu}_{\rm s}}
\newcommand{\nuDBar}{\bar{\nu}_{\rm D}}

\newcommand{\lnLc}{\ln\Lambda_{\rm c}}

\newcommand{\ddint}{\displaystyle\int}

\usepackage[T1]{fontenc}
\usepackage{bm}
\usepackage{color}
\usepackage{graphicx}
\usepackage[breaklinks=true]{hyperref}
\hypersetup{
  unicode=false,          
  pdftoolbar=true,        
  pdfmenubar=true,        
  pdffitwindow=false,     
  pdfstartview={FitH},    
  pdfsubject={Subject},   
  pdfcreator={Creator},   
  pdfproducer={Producer}, 
  pdfkeywords={keyword1} {key2} {key3}, 
  pdfnewwindow=true,      
  colorlinks=true,        
  linkcolor=blue,         
  citecolor=blue,         
  filecolor=blue,         
  urlcolor=blue           
}

\title[Runaway dynamics in ITER-like disruptions]{Runaway dynamics in the DT phase of ITER operations in the presence of massive material injection} 
\author{O.~Vallhagen\corresp{\email{vaoskar@chalmers.se}},   O.~Embreus, I.~Pusztai, L.~Hesslow, T.~F\"ul\"op}
\affiliation{Department of Physics, Chalmers University of Technology,
 SE-41296 G\"{o}teborg, Sweden}

\begin{document}
\maketitle
\begin{abstract}
 A runaway avalanche can result in a conversion of the initial plasma
 current into a relativistic electron beam in high current tokamak
 disruptions.
  We investigate the effect of massive material injection of
  deuterium-noble gas mixtures on the coupled dynamics of runaway
  generation, resistive diffusion of the electric field, and
  temperature evolution during disruptions in the DT phase of ITER operations.
We explore the dynamics over a wide range of injected concentrations
and find substantial runaway currents, unless the current quench time
is intolerably long.
The reason is that the cooling associated with the injected
material leads to high induced electric fields that, in combination
with a significant recombination of hydrogen isotopes, leads to a
large avalanche generation. Balancing Ohmic heating and radiation
losses provides qualitative insights into the dynamics, however, an
accurate modeling of the temperature evolution based on energy balance
appears crucial for quantitative predictions.
\end{abstract}

\maketitle


\section{Introduction}
One of the critical areas of research supporting the successful operation of ITER, and other
large-current tokamaks, is the development of techniques to mitigate the high thermal and magnetic energies released in plasma-terminating disruptions \citep{Boozer2015}. The heat loads resulting from the sudden loss of thermal energy---the thermal quench (TQ)---and the electromechanical stresses induced by the
subsequent loss of the plasma current---the current quench (CQ)---might cause severe damage to plasma facing
components (PFCs) \citep{Lehnen2015}. 

Massive material injection (MMI) has been considered as a possible method to safely terminate
the discharge and avoid disruption related damage \citep{HollmannDMS}.
When injected into a disrupting plasma, impurity atoms radiate away energy from the plasma isotropically, thereby
reducing localized heat loads.  The choice of the impurity, quantity and injection scheme provides means to control the CQ
duration to a certain extent. This is important since the CQ duration determines the magnitude of the induced eddy currents in the wall, and thus the related mechanical forces, together with the toroidal electric field responsible for the generation of runaway electrons (REs).

During the CQ phase of disruptions in reactor-scale devices, such as ITER, large RE currents are expected to form \citep{Boozer2015,Breizman_2019}. These energetic electrons are of particular concern, as they may give rise to localized power deposition and cause melting of PFCs. ITER will have approximately an order of magnitude higher plasma current
compared to current devices, which makes a substantial difference in the potential e-folds of the seed current \citep{RosenbluthPutvinski1997}. As MMI has a major impact on the runaway dynamics, it is also being considered as a runaway mitigation method. However, no clear strategy regarding runaway mitigation with MMI has emerged yet. 

Recent results by \citet{Hesslow_2019} indicate a substantial increase in the avalanche multiplication gain in the presence of massive material injection of heavy impurities
compared to previous estimates. The reason is that in the presence of partially ionized impurities, the increased number of target electrons available for the avalanche
multiplication of runaways is only partially compensated by the increased friction force.

In this paper, we address the runaway current formation in connection with material injection in ITER disruption scenarios. The simulations
are performed with a one-dimensional runaway fluid code, based on the {\sc go} framework, described in Sec.~\ref{go}. {\sc go} has been used in the past  for evaluating material injection scenarios \citep{gal,Feher}, and for interpretative modelling of experiments  \citep{Papp_2013}. The version used in this paper includes Dreicer, tritium
decay and Compton contributions to the runaway seed formation, and avalanche generation.  A careful modelling of the effect of partially ionized atoms is particularly important as we consider heavy noble gas impurities. Thus, we will use the avalanche growth rate derived by \cite{Hesslow_2019}, which has been carefully benchmarked to
kinetic simulations. In Sec.~\ref{mmi}, we will present simulations in ITER scenarios, with neon and argon impurity injections, as well as mixed impurity-deuterium injections.

The model used here contains self-consistent calculations of the
temperature and electric field evolution during the current quench,
including the main runaway generation mechanisms except hot-tail
generation.  The omission of part of the hot-tail source is motivated
with radial losses due to the breakup of magnetic surfaces that
accompanies the thermal quench. Indeed, recent work based on
fluid-kinetic simulations shows that taking into account all the
hot-tail electrons overestimates the final runaway current by
approximately a factor of four in ASDEX Upgrade
\citep{Hoppe2020}. Even if it is difficult to know how large
part of the hot-tail runaways become deconfined in an ITER thermal
quench, it is reasonable to assume that some of the hot-tail seed
survives.  Indeed, deeply trapped electrons do not get lost even when
the magnetic surfaces break up, and can be scattered into the passing
region and run away after the magnetic surfaces have
re-formed. Therefore, the model presented here is likely to
underestimate the runaway seed.

Even in the absence of the hot-tail seed, our results indicate that if
losses due to magnetic perturbations do not occur during a large
fraction of the current quench, impurity injection leads to high
runaway currents in the DT phase of ITER operation, even if it is
combined with deuterium injection.  The reason is that the cooling
associated with a large amount of injected material results in low
temperatures leading to recombination and corresponding high value of
the total-to-free electron density ratio, which in turn, enhances the
avalanche growth rate. A consequence of these results is that
successful runaway mitigation during the non-nuclear phase of ITER
operation may not provide a sufficient validation of the ITER
disruption mitigation strategy, since the presence of radioactive
sources of superthermal electrons during the DT phase changes the
dynamics.

\section{Current dynamics and temperature evolution during material injection}
\label{go}
In a cooling plasma, the conductivity drops and an electric field is
induced to maintain the plasma current. The evolution of the parallel electric field $E_\|$ is modelled by the flux-surface averaged induction
equation \citep{Fulop_2020}
\begin{align} \frac{1+\kappa^{-2}}{2}\frac{1}{r}\frac{\partial }{\partial r}
  r\frac{\partial E_\|}{\partial r}=\mu_0 \frac{\partial j_\parallel}{\partial
    t},
\label{GO}
\end{align}
where $\kappa$ is the elongation (assumed to be constant here), and $j_\parallel=\sigma_{\rm Sp} E_\|+j_\text{RE}\approx \sigma_{\rm Sp} E_\|+ecn_\text{RE}$
is the sum of Ohmic and runaway current densities, with $n_\text{RE}$
being the number density, $j_\text{RE}$ the current density of runaways, and $\sigma_{\rm Sp}$ is the Spitzer conductivity.  Furthermore, $r$ denotes the minor radius at the mid-plane,  while $\mu_0$, $e$, and $c$ denote the magnetic permeability, the elementary charge and the speed of light, respectively. The boundary conditions for equation \eqref{GO}, are $\left. \partial E_\parallel/\partial r\right|_{r=0}=0$, while at $r=a$, where $a=2\,\rm m$ is the mid-plane minor radius of the plasma, $E_\parallel$ is determined by a perfectly conducting wall at $r=b=2.15\,\rm m$. Matching the solution for $r<a$ to the vacuum solution for $a<r<b$ gives $E_\parallel(a)=a\ln{(a/b)}\left. \partial E_\parallel/\partial r\right|_{r=a}$. Note that $j_\|$ and $E_\|$ represent flux-surface averaged values. Toroidicity effects are neglected here.  

When the electric field is larger than a critical field, runaways are produced by velocity space
diffusion into the runaway region due to small angle collisions
(Dreicer generation), and, in deuterium-tritium (DT) operation, by tritium decay and  Compton scattering of
$\gamma$ photons originating from the activated wall. In addition, existing runaways
can create new ones through close collisions with lower energy electrons \citep{RosenbluthPutvinski1997}.  The resulting exponential growth in the number of REs---the avalanche---can be substantially increased in disruptions mitigated by MMI, according to recent estimates by \citet{Hesslow_2019}.

\subsection{Runaway generation rates}
To compute the Dreicer runaway generation rate for given plasma conditions, we use a neural
network\footnote{The neural network is available at
  \href{https://github.com/unnerfelt/dreicer-nn}{https://github.com/unnerfelt/dreicer-nn}}
\citep{HesslowDreicer} trained on a large number of kinetic
simulations by the {\sc code} kinetic equation solver
\citep{CODEPaper2014}, which accounts for the effect of partially ionized atoms as described by \cite{HesslowJPP}.  In ITER-like disruptions we find, however, that Dreicer generation is always negligible compared to the other reactor-relevant seed generation mechanisms, hot-tail generation, tritium decay and Compton scattering, but for completeness it is included in the simulations.

The runaway seed produced by tritium decay is modelled as
\citep{MartinSolis2017,Fulop_2020}
\begin{equation}
\left(\frac{\partial n_\mathrm{RE}}{\partial t}\right)^\mathrm{tritium}=\ln{(2)}\frac{n_\mathrm{T}}{\tau_\mathrm{T}}f\left(W_\mathrm{crit}\right),
\label{tritium}\end{equation}
where $n_\mathrm{T}$ is the tritium density,
$\tau_\mathrm{T}\approx 4500$ days is the half-life of tritium,
$f(W_\mathrm{crit})\approx 1-(35/8)w^{3/2}+(21/4)w^{5/2}-(15/8)w^{7/2}$
is the fraction of the electrons created by tritium decay above the
critical runaway energy $W_\mathrm{crit}$, with $w=W_\mathrm{crit}/Q$ and $Q=18.6\,\rm keV$. The critical runaway energy $W_\mathrm{crit}$ is given by
$W_\mathrm{crit}=m_{\rm e} c^2\left(\sqrt{p_\star^2+1}-1\right)$, in terms of
\begin{equation} p_\star = \nofrac{\sqrt[4]{\nuSBar(p_\star)
      \nuDBar(p_\star)}}{ \sqrt{E_\parallel/\Ec },
  }\label{pstar}\end{equation}
the critical momentum for runaway acceleration, where the runaway probability makes a rapid transition between values of essentially $0$ and $1$, as shown in Appendix~\ref{appA}. We normalize momenta to $m_{\rm e} c$, and we have introduced the critical electric field $E_{\rm c}=n_{\rm e} e^3 \ln \Lambda /(4 \pi \epsilon_0^2 m_{\rm e} c^2)$, where $n_{\rm e}$ is the free electron density,  $ \lnLc \,{\approx}\,14.6+0.5 \ln (T_{\rm eV}/n_{\rm e20})$ is the relativistic Coulomb
logarithm,
$T_{\rm eV}$ is the electron temperature in electronvolts and
$n_{\rm e20}$ is the density of the background electrons in units of
$\unit[10^{20}]{m^{-3}}$, $\epsilon_0$ is the vacuum permittivity, and $m_{\rm e}$ is the electron mass.

Here, the normalized slowing-down and deflection frequencies, $\nuSBar$ and $\nuDBar$, are defined in terms of the full ones ($\nu_{\rm s}$ and $\nu_{\rm D}$) through equations~(5) and (9) of \citep{Ecrit},
\begin{equation}
\nu_{\rm s} = \frac{eE_{\rm c}}{m_{\rm e} c}\frac{\gamma^2}{p^3}\bar\nu_{\rm s}, \qquad 
\nu_{\rm D} =\frac{eE_{\rm c}}{m_{\rm e} c}\frac{\gamma }{p^3}\bar\nu_{\rm D}.
\end{equation}
In the case of a fully-ionized plasma and with a constant Coulomb logarithm, $\bar\nu_{\rm s} = 1$ and $\bar\nu_{\rm D} = 1+Z\sub{eff}$.

Runaways can also be created via Compton scattering of
electrons to the runaway region in momentum space. These events are caused by $\gamma$ photons emitted from the plasma facing components that are activated by
the neutrons produced in the DT fusion reactions. Using radiation transport calculations performed at several poloidal locations in ITER, \cite{MartinSolis2017} estimated the gamma flux energy spectrum  to be
$ \Gamma_\gamma (E_\gamma) =\Gamma _{\gamma 0} \exp{(-\exp{(z)}-z+1)}$, with
$z=\left[\ln{(E_\gamma [{\rm MeV}])}+1.2\right]/0.8 $ and $\Gamma _{\gamma 0}=4.44\cdot 10^{17}\,\rm m^{-2}s^{-1}MeV^{-1}$, giving a total flux of $10^{18}\,\rm m^{-2}s^{-1}$, when integrated over energy.  Using the
expression for the total Compton cross-section \citep{MartinSolis2017}
    \begin{eqnarray}
      \sigma(E_\gamma)= \frac{3 \sigma_{\rm T}}{8}\left\{ \frac{x^2-2x-2}{x^3}\ln{\frac{1+2x}{1+x(1-\cos{\theta_{\rm c}})}}\hspace{5cm}\right.\nonumber\\ \left.+\frac{1}{2x}\left[\frac{1}{\left[1+x(1-\cos{\theta_{\rm c}})\right]^2}-\frac{1}{(1+2x)^2}\right]-\frac{1}{x^3}\left[1-x-\frac{1+2x}{1+x(1-\cos{\theta_{\rm c}})}-x \cos{\theta_{\rm c}}\right]\right\}, 
      \end{eqnarray} 
      with 
      $$ \cos{\theta_{\rm c}}=1-\frac{m_{\rm e} c^2}{E_\gamma} \frac{W_\mathrm{crit}/E_\gamma}{1-(W_\mathrm{crit}/E_\gamma)},$$
the Thomson scattering cross section $\sigma_{\rm T}=8\pi/3[e^2/(4\pi\epsilon_0m_{\rm e} c^2)]^2$, and $x=E_\gamma/(m_{\rm e} c^2)$, the runaway generation rate can be evaluated as
    $$
   \left(\frac{\partial n\sub{RE}}{\partial t}\right)^\gamma =n_{\rm e} \int\Gamma_\gamma (E_\gamma)\sigma (E_\gamma)d E_\gamma.
    $$ Note, that the Compton seed depends on the final configuration
   of the first wall and blanket, as well as the time elapsed after
   the DT reactions cease, therefore the numbers used here are only
   estimates.

Close collisions between existing runaways and thermal electrons generate new runaways, leading to an exponential growth of
the runaway density, from a usually tiny seed population. In partially ionized plasmas, the
growth rate for this avalanche process is influenced by the extent to which fast
electrons can penetrate the bound electron cloud around the impurity ion, i.e. the
effect of partial screening. Taking into account these effects, the growth rate is given by \citep{Hesslow_2019}
\begin{equation}
  \left(\frac{\partial n_\text{RE}}{\partial
      t}\right)_\text{aval}^\text{screened} =    \frac{en_\text{RE}}{m_{\rm e} c \lnLc} \frac{\netot}{\nef} \frac{E_\parallel-\Eceff}{\sqrt{4+
      \nuSBar(p_\star) \nuDBar(p_\star) 
    }},
\label{Hava}
\end{equation}
where $\nef$ and $\netot$ is the free and the total (free+bound) electron densities, respectively~\citep{HesslowJPP}.
The
expression for \Eceff\ was derived in~\citep{Ecrit}\footnote{A
  numerical implementation of \Eceff\ is available at
  \url{https://github.com/hesslow/Eceff}}. To obtain a well-behaved formula also for $E_\parallel< \Eceff$, which we use to approximately describe the runaway decay at near-critical electric fields (neglecting losses due to magnetic perturbations), we replace $p_\star(E_\parallel)$ by $p_\star(\Eceff)$ for $E_\parallel<\Eceff$.

In the completely screened limit, when the electron only interacts with the net charge of the ion, the avalanche growth rate reduces to
\citep{RosenbluthPutvinski1997}
\begin{equation} \left(\frac{\partial n_\text{RE}}{\partial
      t}\right)_\text{aval}^\text{RP}= \frac{e n_\text{RE}}{m_{\rm e} c \lnLc}
  \frac{E_\parallel-\Ec}{\sqrt{5+Z_{\rm eff} }}.
\label{RPava}
\end{equation}

The Dreicer and the avalanche runaway generation processes can be
reduced by finite aspect ratio effects. However, recent results by
\cite{McDevitt_2019} indicate that at high densities and electric
fields this reduction is negligible due to the high collisionality of
electrons at momentum $p_\star$. In such circumstances, the
bounce-averaged approach for collisionless electrons, previously
employed to take into account the effect of toroidicity, is not valid,
and the runaway generation is approximately local. As we consider
cases with massive material injection, leading to high densities, in
this paper the commonly used neoclassical factor that would multiply
the avalanche growth rate
[$\varphi_\epsilon=(1+1.46\epsilon^{1/2}+1.72 \epsilon)^{-1/2}$, with
$\epsilon$ the inverse aspect ratio] will be omitted.

    Hot-tail generation occurs in the case of sudden cooling, when the
    collision frequency is lower than the cooling rate, and fast
    electrons do not have time to thermalize. Hot-tail generation is
    predicted to be the dominant primary generation when the TQ
    duration is shorter than the collision time at the runaway
    threshold velocity \citep{helanderhottail}, which is likely to be
    the case in ITER \citep{SmithHottail2005}. However, hot-tail
    runaways are produced in the early phase of the TQ when the level
    of magnetic fluctuations is high, and their losses due to radial
    transport is likely to be significant.  Lacking reliable models
    for self-consistently treating the hot-tail generation and the
    losses due to magnetic fluctuations during the TQ, both of these
    processes will be ignored in this paper. The implications of a remnant hot-tail seed will be discussed in Section~\ref{sec:discussion}.

\subsection{Temperature evolution} The major causes of energy loss during the disruption are radial transport due to magnetic fluctuations, induced by magnetohydrodynamic (MHD) instabilities, and line radiation due to impurity influx. The MHD-induced energy loss is likely to dominate in the initial part of the thermal quench until the temperature has dropped to $\sim 100$ eV, due to its strong temperature scaling $\sim T^{5/2}$ \citep{Ward_1992}. For simplicity, this part of the temperature drop is modelled by an exponential decay according to
\begin{equation}
    T(r,t)=T_{\rm f}(r)+[T_0(r)-T_{\rm f}(r)]e^{-t/t_0},
    \label{eq:expdecay}
\end{equation}
where the temperature decays from the initial $T_0(r)$ towards the final $T_{\rm f}(r)$ temperature profile with a characteristic time constant $t_0$. This temperature evolution is employed until the central temperature drops to $\sim 100\,\rm eV$. After this, the MHD-induced losses are assumed to be negligible, and the temperature evolution is determined by the energy balance equation
\begin{eqnarray}
    \frac{3}{2}\frac{\partial  (n T)}{\partial t}&=\displaystyle\frac{1+\kappa^{-2}}{2}\frac{3
      n}{2 r}\frac{\partial
    }{\partial r}\left(\chi r \frac{\partial T}{\partial
        r}\right)+\sigma(T, Z_{\rm eff})E^2-\sum_{i,k}n_\mathrm{e}n_{k}^iL_{k}^i(T,n_{\rm e})-P_{\rm Br}-P_{\rm ion}.     
    \label{eq:ebalance}
\end{eqnarray}
where $n$ is the total density of all species (electrons and ions),
$n^k_{i}$ is the density of the $i^{\rm th}$ charge state
($i=0, 1, ..., Z-1$) of the ion species $k$ (e.g.~deuterium, neon),
$P_{\rm Br}[\rm W m^{-3}]= 1.69\cdot 10^{-38} (n_{\rm e}[\rm m^{-3}])^2 \sqrt{T[\rm eV]} Z_{\rm eff}$ is the
Bremsstrahlung radiation loss and $P_{\rm ion}=\sum_{i,k} E_k^i I_k^i n_k^i n_{\rm e}$ is the ionization energy
loss. Here, $I_k^{i}$ denotes the electron impact ionization rate and $R_k^i$ 
the radiative recombination rate for the $i^{\rm th}$ charge state of
species $k$, respectively, and $E_k^i$ denotes the corresponding ionization energy. The ionization and recombination rates, as well as the line radiation rates $L_k^{i}(n_{e},T)$, are extracted from the Atomic Data and Analysis Structure (ADAS) database\footnote{ADAS: \url{http://www.adas.ac.uk}}. The heat diffusion term in the equation is included for
completeness, but its effect is negligible in all the considered
cases.  In the simulations we use the constant heat diffusion
coefficient $\chi=1\,\rm m^2/s$, but  values in the
range $0.1$ to $100\,\rm  m^2/s$ give similar final runaway currents.

We calculate the density of each charge state for every ion species from the time dependent rate equations
\begin{equation}
    \frac{\ud n_{k}^i}{\ud t}=  n_{\rm e} \left[I_k^{i-1} n_{k}^{i-1} - (I_k^i+ R_k^i) n_{k}^{i} +  R_k^{i+1} n_{k}^{i+1} \right].
 \label{eq:tdre}\end{equation}

To allow for comparison to previous work by \cite{MartinSolis2017}, we will also show results using a temperature evolution assuming equilibrium between Ohmic heating and line radiation losses, so that the temperature profile satisfies 
\begin{equation}    
    E^2\sigma[T,Z_\mathrm{eff}(T)]=\sum_i n_\mathrm{e}(T)n^i_k L^i_k [T,n_{\rm e}(T)].     
    \label{eq:rad_equilibrium}   
  \end{equation}
Equation \eqref{eq:rad_equilibrium} is solved numerically using $T=5$ eV as initial temperature. In this case, the densities of the various ionization states, and the corresponding electron density, are calculated assuming an equilibrium between ionization and recombination; that is, $n^{i}_k$ are computed from
\begin{equation}
   \begin{split}
    &R^{i+1}_k n^{i+1}_k-I^i_k n^i_k=0,  \hspace{0.5cm}i=0,1,...,Z-1,\\
    &\sum_i n^{i}_k=n^\mathrm{tot}_k,  \hspace{0.5cm}i=0,1,...Z,
    \end{split}
    \label{eq:ioniz_equilibrium}
\end{equation}
where $n^\mathrm{tot}_k$ is the total density of species $k$.

\section{Effect of massive material injection}
\label{mmi}
In the following we present simulations of an ITER-like current quench
with material injection. For the scenario we consider, the initial
plasma current is $I_{\rm p}(t=0)=15\,\rm MA$, and the major and minor
radii are $R=6\,\rm m$ and $a=2\,\rm m$, respectively. The
pre-disruption density profile is assumed to be flat with a
value of $n_\text{e}(t=0)=10^{20}\,\rm m^{-3}$. The initial
temperature profile is given by $T(t=0,r) = 20 [1-(r/a)^2] \,\rm keV$.
The initial current density profile is assumed to be
$j_\|(t=0,r)=j_\text{0} [1-(r/a)^{0.41}]$, with the 
normalization parameter $j_\text{0}$ chosen to give a total plasma current of $15\;\rm MA$ (for a non-elongated plasma, $\kappa=1$, it is $j_\text{0}=1.69 \,\rm MA/m^2$). The current density profile $j_\|(t=0,r)$ corresponds to an
internal inductance of $l_\text{i}=0.7$. This set of initial plasma parameters is similar to that used by \citet{MartinSolis2017}.

We solve the induction equation, (\ref{GO}), with the runaway
generation rate given by the sum of the primary (Dreicer+tritium
decay+Compton) and avalanche growth rates, for different
amounts of impurity and deuterium. We find that the tritium or Compton seed
dominates over Dreicer in all the considered cases.

The seed from tritium decay decreases with increasing impurity content. This is partly due
to the shorter current quench times associated with higher impurity
content, leaving the seed less time to be generated, and partly because
the critical energy, $W_{\rm crit}$, increases with increasing impurity content, so
that a smaller fraction of the tritium spectrum falls within the
runaway region.

On the other hand, the Compton seed increases with increasing impurity
content, due to the increasing number of available target electrons
for Compton scattering. The energy of the $\gamma$ photons is much
larger than the ionization potential for all species present in the
plasma, hence both bound and free electrons can become runaways due to
Compton scattering. Furthermore, the energy of the $\gamma$ photons is
much larger than the critical runaway energy, so an increase in the
critical runaway energy only has a marginal effect on the runaway
generation due to Compton scattering by preventing electrons scattered
at large angles from becoming runaways.

As a result, a seed current of the order of a few amperes is obtained
almost independently of the injected amount of noble gas and/or
deuterium.   As we will see, even if the seed current is very small, it
results in a large final runaway current, due to the substantial
avalanche effect. 

We assume the injected material to be uniformly distributed at the
beginning of the simulation. For the initial part of the thermal
quench, we use an exponentially decaying temperature evolution
according to \eqref{eq:expdecay} with $t_0=1$ ms and $T_{\rm f}(r)=50$ (flat
profile), until $t=6$ ms, when the central temperature has dropped to $\approx 100
\,\rm eV$. This represents the initial phase of the disruption when the
MHD-induced energy loss dominates. Below $100 \,\rm eV$ the temperature is
determined by the energy balance equation \eqref{eq:ebalance}.  The
density of the charge states for every ion species is calculated from
the time dependent rate equations \eqref{eq:tdre} during the whole
simulation, both in the initial exponential decay phase and when the
energy balance equation is employed. The main reason for using the
exponentially decaying temperature in the initial phase is to
determine the initial charge states of the impurities when the
temperature reaches $100 \,\rm eV$, and the energy balance equation is
invoked. We have confirmed that the results are insensitive to the value of
the decay time, and the value of the heat diffusion coefficient
$\chi$. 

\subsection{Pure neon injection}
\begin{figure}
  \centering
\includegraphics[width=0.96\textwidth]{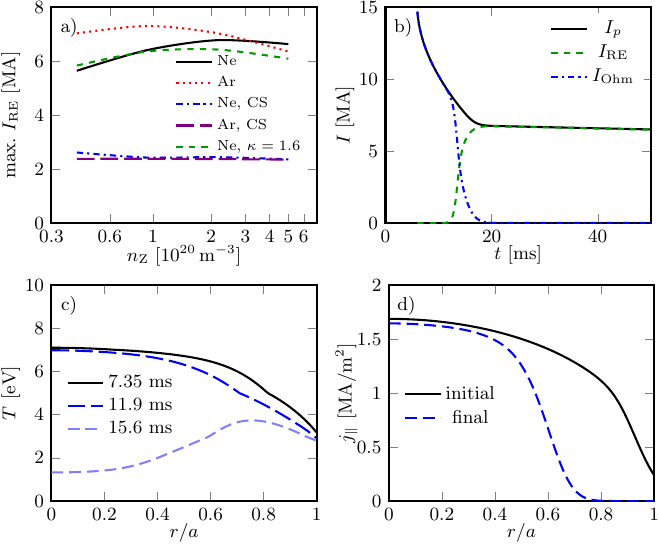}  
  \caption{a) Maximum runaway current as a function of injected noble gas density. Solid: neon with partial screening (i.e.~avalanche calculated with \eqref{Hava}); dash-dotted: neon complete screening ("CS", i.e.~using \eqref{RPava}); dotted: argon with partial screening; long dashed: argon complete screening; dashed: neon with screening at $\kappa=1.6$.  Panels~b)-d) correspond to $10^{20}\,\rm m^{-3}$ neon injection, with screening effects included and $\kappa=1$. b) Temporal evolution of plasma current (solid), and breakdown into Ohmic (dash-dotted) and runaway (dashed) contributions. c) Snapshots from the time evolution of the temperature. d) Initial current profile (solid) and runaway current profile at the end of the Ohmic CQ (dashed).\label{Case1} }
\end{figure}
To illustrate the effect of impurity injection, we start by simulating the runaway dynamics for pure noble-gas injections.  Figure \ref{Case1}a shows the runaway current as a function of injected neon and argon density (normalized to the initial deuterium density, $10^{20}\,\rm m^{-3}$),  using two models for avalanche generation: with partial screening effects, as given in equation \eqref{Hava}, and with complete screening (\ref{RPava}), respectively. The effect of screening is substantial.  It increases the final runaway current from about $2$-$3\,\rm MA$ without screening to $6$-$7\,\rm MA $ with screening, for both neon and argon injections. Including partial screening, i.e.~using \eqref{Hava}, the avalanche growth rate increases with density for neon injection, while in the completely screened case, i.e.~using \eqref{RPava}, it is slightly decreasing.  Including the effect of the elongation reduces the final runaway current at higher injected neon  densities, but only marginally.

Figures \ref{Case1}b-d show  details of the current and temperature evolution during the CQ for a representative case with $n_{\rm Ne}=10^{20}$ m$^{-3}$. The temperature stabilises at a few eV at the centre with a rather flat radial profile, resulting in a CQ time scale of the order of a few tens of milliseconds. The CQ is completed at about 20 ms (14 ms after the TQ) and results in the  formation of a runaway beam with a current of 6.7 MA.  At this time the radial profile of the runaway beam is slightly more peaked around the magnetic axis compared to the initial current profile. The dissipation rate (after the 20 ms) is very slow for this relatively modest injected impurity density.

\subsection{Mixed deuterium and neon injection}

\begin{figure}
\includegraphics[width=0.96\textwidth]{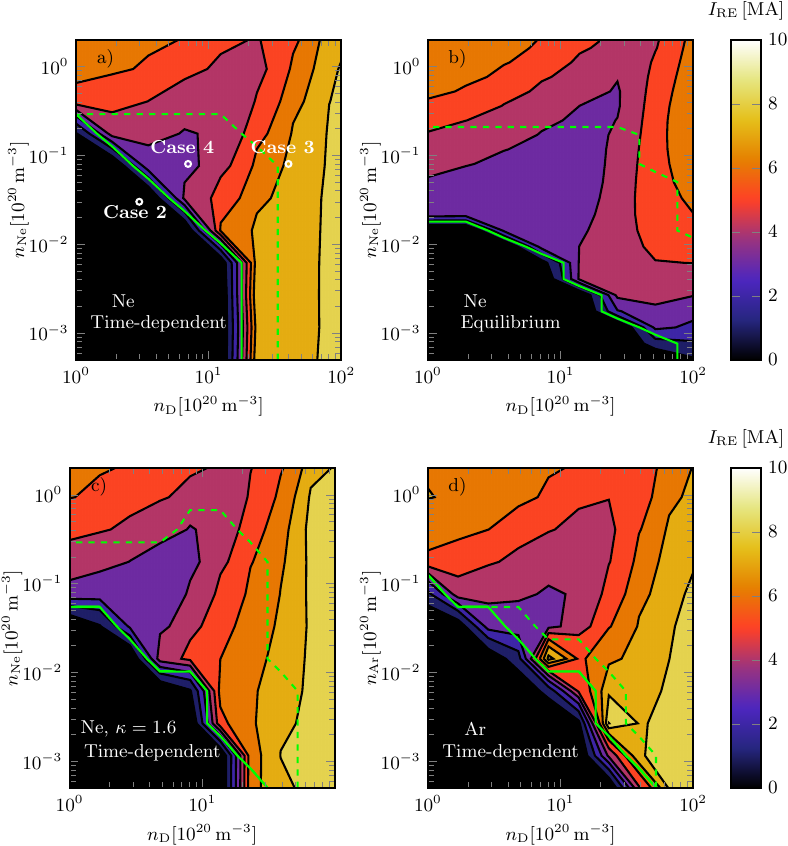} 
\caption{Maximum runaway current as a function of injected density
     for the reference ITER-like scenario. The horizontal and vertical
     axes show the injected deuterium and noble gas
     densities (neon in a-c, argon in d), respectively.  The
     temperature evolution is determined by the equations
     \eqref{eq:expdecay} and (\ref{eq:ebalance}) in (a), and
     \eqref{eq:rad_equilibrium} in (b). c) As in (a) but with
     plasma elongation $\kappa=1.6$. d) As in (a) but for
     argon. Below the green solid line the Ohmic CQ time
is longer than $150\,\rm ms$ and above the green dashed line the Ohmic CQ time
is shorter than $35\,\rm ms$. }
  \label{Ne}
\end{figure}

We now turn to simulating the effect of injections of mixtures of
noble-gas and deuterium using the same approach as in the previous
section. Runaway currents right before the dissipation phase
(i.e.~when $I_{\rm RE}$ assumes its maximum) are shown in
figure~\ref{Ne}, for different amounts of injected noble gas and
deuterium. Below the green solid line the injected gas mixture is
insufficient to induce a complete radiative collapse, which we
characterise by requiring that the CQ time, defined as $t_{\rm CQ}=[t(I_{\rm Ohm}=0.2I_{\rm p}^{(t=0)})-t(I_{\rm Ohm}=0.8I_{\rm p}^{(t=0)} )]/0.6$, is longer than $150\,\rm
ms$. For these injection parameters, the temperature remains of the
order of $100\,\rm eV$ in parts of the plasma, and the corresponding
CQ times in this region are therefore very long (on the order of
seconds). Above the green dashed line, the Ohmic current quench time is
less than $35\,\rm ms$, which is the boundary to avoid damage due to torques
on the first wall \citep{HollmannDMS}. Note, however, that in cases with a large runaway conversion, the runaway current aborts the current quench rather abruptly, so that the ohmic CQ time calculated here is a lower estimate of the CQ time in the absence of runaways. In the region between the solid and dashed green lines, we obtain
runaway currents ranging from $3$ to $8\,\rm MA$.

Comparing the case with time-dependent temperature, equations
\eqref{eq:expdecay} and (\ref{eq:ebalance}), with the case when an
equilibrium between Ohmic heating and line radiation is assumed,
equation \eqref{eq:rad_equilibrium}, we find that the former leads to
higher runaway currents, especially for high deuterium contents,
cf.~figures~\ref{Ne}a and \ref{Ne}b.  One reason for this is that in
the time-dependent case, it takes a finite time to reach the
equilibrium ionization distribution. At low degrees of ionization,
both higher ionization states of neon and the corresponding higher
electron density leads to larger radiative losses, which decreases the
temperature. This, in turn, leads to higher induced electric fields,
and thus higher runaway currents. Also, if the temperature drops below
$\approx 2\,\rm eV$, the deuterium starts to recombine. The
corresponding increase in partially ionized species enhances the
avalanche, which also leads to higher runaway currents, as will be
discussed more in detail later.

Another reason for the difference is
that ionization energy losses are included in the time-dependent
approach, which further lowers the temperature.  Note, that ionization
losses are operational even when there is no net increase in free
electron density (and thus there is no net increase in the chemical potential of ionized species), as radiative recombination may balance or outweigh
the collisional ionization events that take place continuously.  As a possible intermediate step between the time-dependent energy balance, (\ref{eq:ebalance}), and 
the Ohmic-radiative equilibrium, \eqref{eq:rad_equilibrium}, one may also consider evolving the temperature according to Eq.~(11) of \citep{Aleynikov2017}, where the line radiation and ionization coefficients assume the ionization states to be in a coronal equilibrium and radiative recombination is disregarded, but the heat capacity and chemical potential of ionized species are included. This would lead to results similar to our figure~\ref{Ne}b, indicating that accounting for the heat capacity and chemical potentials of the ionized species does
not make a major difference. 

Finally, starting the iterative solution of \eqref{eq:rad_equilibrium}
from an initial temperature of $5\,\rm eV$ ensures that a rather low
equilibrium temperature is obtained whenever that exists, while the
time-dependent approach can evolve towards a higher equilibrium
temperature.  This mainly affects position of the solid green line in figure
\ref{Ne}.

The results are similar for elongated plasmas, see figure~\ref{Ne}c
for a radially constant $\kappa=1.6$. Plasma elongation generally reduces the runaway
generation due to its significant effect on Dreicer generation \citep{Fulop_2020}, but it
has only a marginal effect for this ITER-like scenario, where tritium decay
and Compton seed dominate over Dreicer generation. Elongation does, however, extend the parameter regime of interest as constrained by the CQ times. Injecting argon instead of neon leads to marginally higher runaway
currents for certain parameters (and reduces the parameter regime of interest as constrained by the CQ times), compare figure \ref{Ne}a with
\ref{Ne}d, but the general conclusions are the same.

         \begin{table}
            \centering 
            \begin{tabular}{cccc}
                     Case& $n_{\rm D}/n_{\rm D0}$&$n_{\rm Ne}/n_{\rm D0}$ &$I_{\rm RE}$ [MA]\\
              1&0&1& 6.7 \\
              2&3&0.03&  0 \\
              3&40&0.08&7.3 \\
                                          4&7&0.08&3.7\\
            \end{tabular}
            \caption{Injected material in the four representative cases studied here (three of them are indicated in Fig.~2a). The initial deuterium density is $n_{\rm D0}=10^{20}\,\rm m^{-3}$. The final column shows the runaway currents right before the dissipation phase (i.e.~when $I_{\rm RE}$ assumes its maximum).}
            \label{cases}
          \end{table}

We identify four qualitatively different regions: (1) a region with large conversion at high neon densities and low deuterium densities, (2) a region with very long CQ times and negligible runaway generation, (3) a region with large runaway conversion at high deuterium densities and (4) a region between (1) and (3) with the lowest runaway conversions.
A representative case from region (1) has already been presented in the previous subsection. In what follows, we analyze the  representative cases from the three remaining regions. 
These cases are marked in figure \ref{Ne}a and given in Table~\ref{cases}.

\begin{figure}
\centering
\includegraphics[width=0.99\textwidth]{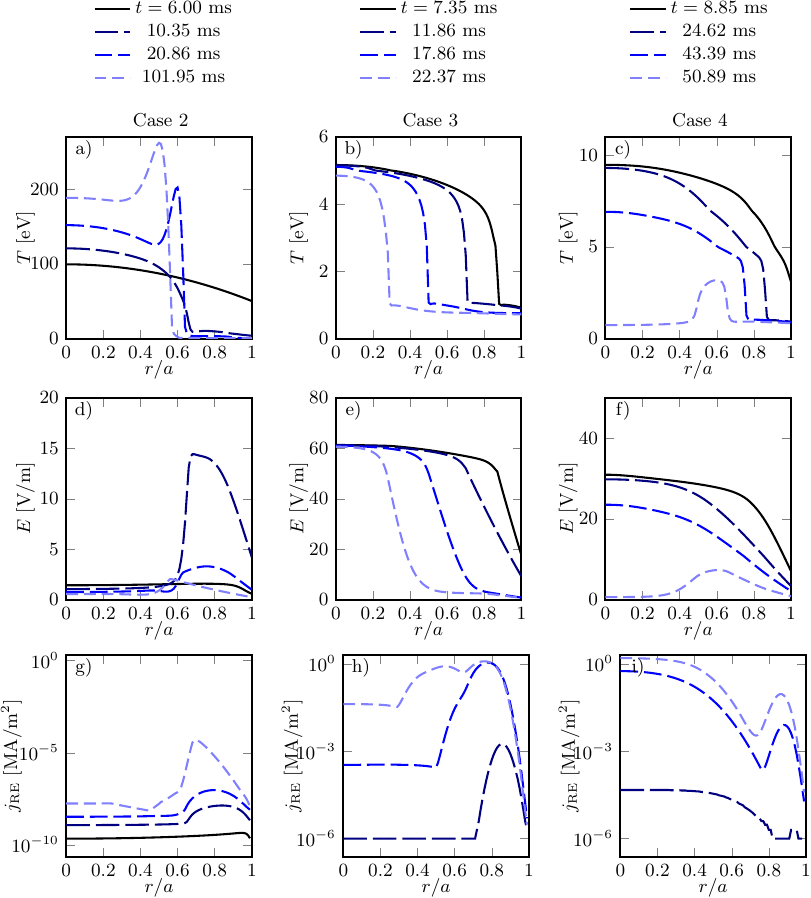} 
\caption{Time slices from the temperature a-c), the electric field d-f) and the runaway current density g-i) evolution. Left column (Case 2):  $n_{\rm Ne}=3\times 10^{18}\,\rm m^{-3}$, $n_{\rm D}=3\times 10^{20}\,\rm m^{-3}$ (very long CQ-time). Middle column (Case 3): $n_{\rm Ne}=8\times 10^{18}\,\rm m^{-3}$, $n_{\rm D}=4\times 10^{21}\,\rm m^{-3}$ (low temperature and high runaway conversion). Right column (Case 4): $n_{\rm Ne}=8\times 10^{18}\,\rm m^{-3}$, $n_{\rm D}=7\times 10^{20}\,\rm m^{-3}$ (moderate runaway conversion).\label{TempCas234}} 
\end{figure}

\begin{figure}
\centering
\includegraphics[width=0.96\textwidth]{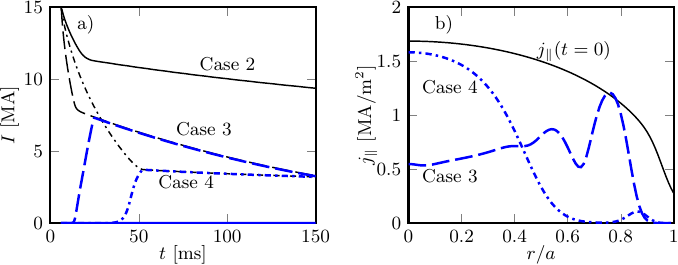}
\caption{a) Current evolution. Thin black: $I_{\rm p}$, thick blue: $I_{\rm RE}$. Solid:  Case 2,  $n_{\rm Ne}=3\times 10^{18}\,\rm m^{-3}$, $n_{\rm D}=3\times 10^{20}\,\rm m^{-3}$ (too long CQ-time); dashed: Case 3, $n_{\rm Ne}=8\times 10^{18}\,\rm m^{-3}$, $n_{\rm D}=4\times 10^{21}\,\rm m^{-3}$ (low temperature and high runaway conversion); dash-dotted: Case 4, $n_{\rm Ne}=8\times 10^{18}\,\rm m^{-3}$, $n_{\rm D}=7\times 10^{20}\,\rm m^{-3}$ (lowest runaway fraction). b) Current density profiles. Thin black line is $j_\|(t=0)$, thick blue lines represent $j_{\rm RE}$ at the time of highest $I_{\rm RE}$, dashed: Case 3; dash-dotted: Case 4.  \label{Case234}}
\end{figure}

The radial profiles of the temperature, electric field and runaway current density at a few time slices are shown in figure \ref{TempCas234} for Cases 2-4. In Case 2, shown in figure \ref{TempCas234}a,d and g, the temperature remains of the order of $100 \,\rm eV$ in the  central part of the plasma, resulting in very long CQ times.  The increasing temperature in the central part of the plasma occurs because the injected material does not cause sufficient radiative losses to counteract the Ohmic heating there. Furthermore, due to the local temperature drop in the edge plasma, a strong electric field is induced (see figure \ref{TempCas234}d), and that will diffuse inward and lead to additional Ohmic heating.  The effect of this increased heating is strongest close to the boundary between the hot and cold regions, where the electric field is induced, resulting in a radial peak also in the temperature. 
The resulting current evolution is shown as the solid curve in figure \ref{Case234}a. After a rather fast drop initially while the current in the cold region decays, the CQ time for the remaining Ohmic current in the hot region is very slow. Notably, the runaway current remains negligibly small throughout the entire process.   

In Case 3 (high injected deuterium density), on the other hand, the radiative losses are strong, resulting in very low temperatures, which leads to a large runaway current, and a full current conversion. The temperature evolution for this case is presented in figure \ref{TempCas234}b. The plasma is divided in two regions: an inner, continuously shrinking region with a temperature of about $5\,\rm eV$, and an outer region with a temperature as low as about $1\,\rm eV$. The boundary between these two regions moves radially inward as the electric field decays away, and the heating decreases (compare figure \ref{TempCas234}b and \ref{TempCas234}e). In this case, however, radiative losses are strong enough to maintain a low temperature in the outer region, even when the electric field is strong enough to cause a significant runaway avalanche in part of the cold region. 

\begin{figure}
  \centering
  \includegraphics[width=0.96\textwidth]{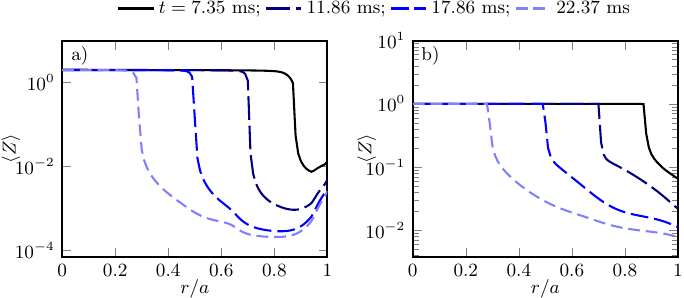}
  \caption{Average ion charge as function of radius  for (a) neon and (b) deuterium for Case 3 ($n_{\rm Ne}=8\times 10^{18}\,\rm m^{-3}$, $n_{\rm D}=4\times 10^{21}\,\rm m^{-3}$). }
  \label{ionization}
\end{figure}

At $1\,\rm eV$, the ionization degree of deuterium is significantly affected, so that a sizable fraction of the deuterium becomes neutral in the outer region of the plasma, as shown in figure \ref{ionization}. This strongly enhances the avalanche, due to the reduction in the free electron density. This enhancement makes the avalanche generation stronger in the outer part of the plasma, even if the electric field is lower there compared to the inner part. The result is a large runaway current of about $7.3\,\rm MA$, with an off-axis maximum, the evolution of which is shown in figure \ref{TempCas234}h, and the final current density profile is shown by the dashed line in figure \ref{Case234}b. There is, however, a significant current dissipation due to the large effective critical electric field, such that by the end of the $150\,\rm ms$ long simulation the remaining runaway current is only $3.2\,\rm MA$.

Finally, Case 4, representing an intermediate case between Case 1 and Case 3, has the lowest conversion, while also an acceptable CQ time. As can be seen in figure \ref{TempCas234}c, the deuterium density is not high enough to result in temperatures low enough to make the deuterium recombine, except close to the edge where the heating and the electric field is very low, but it is sufficiently high to dampen the avalanche, at least partially. The resulting runaway current evolution is shown in figure \ref{TempCas234}i and the total current evolution is shown by the dash-dotted line in figure \ref{Case234}a. The final runaway current is about $3.7\,\rm MA$ with a radial profile centred near the magnetic axis (see figure \ref{Case234}b). The dissipation rate is relatively small with about 13\% dissipation within the simulation time of $150\,\rm ms$.  

A limitation of the model used in this paper is that transport of
neutral particles is not taken into account. Since the neutral
particles are not confined by the magnetic field, they might be lost
before giving rise to a significant enhancement of the
avalanche. However, if a large fraction of the injected deuterium would
recombine and leave the plasma, its contribution to the increase in
the critical electric field would also be lost. Therefore the resulting
runaway current turns out to be similar to a case with a lower amount
of injected deuterium. Simulations, where neutral particles are instantaneously removed from the system, show that the runaway current for
injected densities in the region around Case 3 is still as large as $3$-$4\,\rm MA$,
only a few percent of which is dissipated within the simulation time of $150\,\rm 
ms$.

\subsection{Qualitative analysis}
\label{qa}

\begin{figure}
  \centering
  \includegraphics[width=0.96\textwidth]{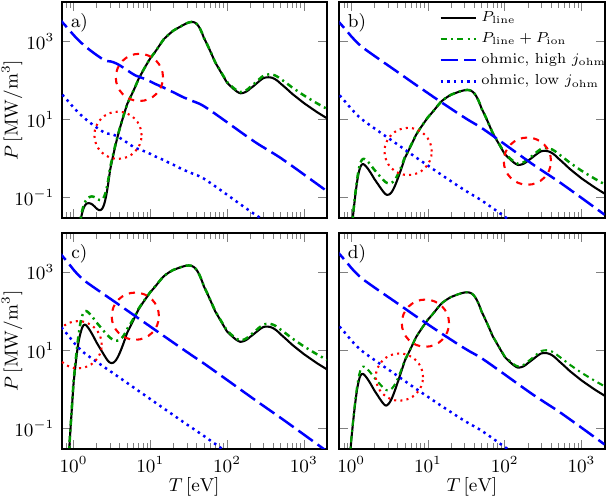}
  \caption{Radiative losses (solid), radiative and ionization losses (dash-dotted)  and Ohmic heating (dashed and dotted) as functions of temperature for Case 1 (a), 2 (b), 3 (c), and 4 (d); assuming the equilibrium distribution over charge states. The Ohmic heating is shown for $j_\mathrm{ohm}=1.69\,\rm MA/m^2$ (dashed) and $j_\mathrm{ohm}=0.2\,\rm MA/m^2$ (dotted); the corresponding stable equilibrium points are marked with circles.}
  \label{Ebalance_illustration}
\end{figure}

The dynamics described in the four cases can be qualitatively understood by considering the behaviour of the avalanche growth rate, together with the balance between radiative losses and Ohmic heating, while assuming an equilibrium distribution over charge states. In figure \ref{Ebalance_illustration}, the radiative losses, the sum of radiative and ionization losses and the Ohmic heating are shown as functions of temperature, for the neon and deuterium densities of the four cases presented above. The Ohmic heating is shown for $j_\mathrm{ohm}=1.69\,\rm  MA/m^2$, corresponding to the initial on-axis current density, and for $j_\mathrm{ohm}=0.2\,\rm MA/m^2$, representing a case where the Ohmic current has partially, but not completely, decayed.

In Case 1, shown in figure \ref{Ebalance_illustration}a, the equilibrium temperatures for both  current densities are on the order of a few $\rm eV$. The avalanche growth rate is enhanced by the presence of a relatively large density of partially ionized neon. Therefore, this case results in a large RE conversion. 

In Case 2, on the other hand, there is an equilibrium temperature at about $200\,\rm eV$ when $j_\mathrm{ohm}=1.69\,\rm  MA/m^2$, as shown in figure \ref{Ebalance_illustration}b. There is also an equilibrium at a lower temperature for this current density. However, since the Ohmic heating is stronger than the radiation losses at the central temperature of about $100 \,\rm eV$ in the beginning of the simulation, where the losses are dominated by radiation, the temperature will increase towards the higher equilibrium. This mechanism causes the inner part of the plasma to remain hot, as observed in figure \ref{TempCas234}a, and is responsible for the very long CQ time. Due to the high temperature, and thus high conductivity, the induced electric field is weak, and the corresponding avalanche growth rate is practically zero, thus RE conversion remains negligible.

With $j_\mathrm{ohm}=0.2 \,\rm MA/m^2$, the (unique) equilibrium temperature shown in figure \ref{Ebalance_illustration}b is still of the order of a few $\rm eV$, corresponding to the cold, outer part of the plasma shown in figure \ref{TempCas234}a. However, the induced electric field remains modest and the low amount of partially ionized material makes the avalanche growth rate---for a given electric field---small, which results in a small runaway generation rate even in this region.

In Case 3, figure \ref{Ebalance_illustration}c shows that the large amount of deuterium leads to an overall enhancement of the radiative losses. For $j_\mathrm{ohm}=0.2\,\rm MA/m^2$, this shifts the equilibrium temperature from a few $\rm eV$ down to only about $1\,\rm eV$, which corresponds to the cold outer part of the plasma in figure \ref{TempCas234}c. This large reduction in the temperature makes a significant difference, as at this temperature, the equilibrium degree of ionization of deuterium is only a few percent. The presence of neutral deuterium enhances the avalanche growth rate, and the low temperature also favours an increased induced electric field. 
These circumstances result in a large avalanche generation in the outer part of the plasma in Case 3, even if the avalanche is partly constrained by a short CQ time and saturation effects. Note, however, that although the Ohmic current density is modest in this region, the Ohmic current remaining in the more central part of the plasma will have to pass through the cold region as it diffuses outwards. Therefore, a significant induced electric field is sustained in the cold region, and thus the local conversion is not limited by the local current density. The result is a large runaway current with an off-axis radial profile, as shown by the dashed line in figure \ref{Case234}b.

Finally, Case 4 can be regarded as a compromise between the various features in the other three cases. The injected densities are sufficiently large to avoid an equilibrium at high temperatures (over $100\,\rm eV$), but not too large, such that equilibrium temperature does not drop too close to $1\,\rm eV$ (unless the Ohmic current density is very low). This ensures an acceptably short CQ time, while still not leading to a large induced electric field. The ratio of the fully ionized deuterium and the partially ionized neon is large enough to limit partial screening effects on the avalanche growth rate, which contributes to  the resulting runaway current remaining modest.

\section{Discussion}
\label{sec:discussion}
When partially ionized impurities are introduced into the plasma, there are two competing effects which affect the avalanche growth rate: (1) additional target electrons become available for avalanche multiplication (represented by the prefactor $n_{\rm e}^\text{tot}$ in the growth rate formula \eqref{Hava}), which leads to an increasing growth rate; (2) the
critical runaway momentum $p_\star$ increases due to the enhancement of the collisional slowing-down  and pitch-angle scattering processes (i.e., through increasing $\nu_{\rm s}$ and $\nu_{\rm D}$, respectively), which leads to a decreasing growth rate. For the scenarios considered here, we  generally find that the additional-target effect prevails, leading to a net increased growth
rate in the presence of impurity injection.

Our results predict a higher runaway conversion  in the presence of massive material injection than previous estimates by \cite{MartinSolis2017}. One of the reasons for the difference is that the avalanche growth rate is different in the two cases, mainly because the corrections to  the slowing-down and deflection frequencies, $\nu_{\rm s}$ and $\nu_{\rm D}$, due to partial screening are different; namely, \citet{MartinSolis2017} employs a fully classical model \citep{Mosher}, whereas we take a quantum-mechanical approach to determine the collision rates. Also, the expressions for the avalanche growth rates are slightly different. The growth rate used here, given in \eqref{Hava}, is based on the calculation by \cite{Hesslow_2019}, where analytical solutions of the kinetic equation were derived in the same parameter regimes as the Rosenbluth--Putvinski calculation, and benchmarked to kinetic simulations. In contrast, \cite{MS2015} considers the expression for the growth rate which is obtained in the momentum-diffusion-free limit, given in equation~(21) of \citep{MS2015} as
\begin{align}
  \Gamma \equiv \frac{1}{ n_\text{RE}}\left(\frac{\partial n_\text{RE}}{\partial
      t}\right)_\text{aval}=\frac{2\pi r_0^2 n_{\rm e}^{\text{tot}} c }{\sqrt{1+(p_\star^\text{MS}/m_{\rm e}c)^2}-1},
  \label{MSava}
\end{align}
where $r_0$ is the classical electron radius, which would agree with our expression in the non-relativistic limit $p_\star \ll m_e c$, if  $p_\star^\text{MS}$ had been defined in the same way. 
Although it is not presented in this form, equation (16) of  \cite{MS2015} defines the effective critical momentum as
\begin{align}
  eE_\parallel = \sqrt{p_\star^2\nu_{\rm s}^\text{MS}(p_\star)[\nu_{\rm D}^\text{MS}(p_\star)+\nu_{\rm s}^\text{MS}(p_\star)]},
  \label{pstarMS}
\end{align}
which is similar to our result (\ref{pstar}), but differs by the appearance
of an additional $\nu_s$ term and by the use of different collision frequencies. The choice of critical momentum
leads to minor differences in the tritium seed current, see
Table~\ref{comptab}.  Our simulations show that the main difference
between the results is not due to the difference between \eqref{MSava}
and \eqref{Hava}, but rather due to the choice of model for $\nu_{\rm s}$ and
$\nu_{\rm D}$.

The other major difference is that here, the spatio-temporal evolution of the temperature is taken into account, while it was assumed to be constant in \citep{MartinSolis2017}. As we have seen in the previous section, the evolution of the temperature has a crucial impact on the runaway dynamics. To quantify the differences, we have performed simulations for three different combinations of argon and deuterium injection, and compared the maximum runaway current values for various levels of sophistication of the modelling; the results are presented in
Table~\ref{comptab}.

When we use the same avalanche growth rate as \citet{MartinSolis2017}, and a constant temperature, we find agreement with their results. In particular, at sufficiently high deuterium injection the generation of runaways is completely suppressed.  However, if the temperature evolution is included, we find a significant runaway conversion. At a fixed value of $n_{\rm Ar}$, the runaway conversion even increases upon an increasing amount of injected deuterium, when using \eqref{Hava}. Furthermore, at high deuterium densities, determining the temperature evolution from the more accurate \eqref{eq:ebalance}, rather than  \eqref{eq:rad_equilibrium}, leads to a further increase in the predicted runaway currents. Note, that the radiation peak resulting from deuterium between 1 eV and 2 eV, that can be seen in figure  \ref{Ebalance_illustration}c has not been taken into account in previous work.  
          \begin{table}
            \centering \small
            \begin{tabular}{cccccccc}
              $n_{\rm D}$&$n_{\rm Ar}$&$I_{\rm seed}$ [A]&$I_{\rm seed}$ [A]&$I_{\rm RE}$ [MA]  &$I_{\rm RE}$ [MA] &$I_{\rm RE}$ [MA] &$I_{\rm RE}$  [MA]\\
              &&$p_\star$ from \eqref{pstarMS} &$p_\star$ from \eqref{pstar}&$\Gamma$ from \eqref{MSava}& $\Gamma$ from \eqref{Hava}& $\Gamma$ from \eqref{Hava}& $\Gamma$ from \eqref{Hava}\\
              &&$T$ const& $T$ const&$T$ const&$T$ const&$T$ from \eqref{eq:rad_equilibrium}&$T$ from \eqref{eq:ebalance}\\

              0&$10^{19}$&0.85& 1.0&3.6   &4.8&6.1 &4.2 \\

              $2\cdot 10^{21}$&$10^{18}$&$8.50\times 10^{-5}$ &0.047 &1.4   &2.3& 3.9 &6.5 \\

              $4\cdot 10^{21}$&$10^{18}$&$2.14\times 10^{-6}$&$5.28\times 10^{-5}$ &0   &0.3&4.2 &6.8 
            \end{tabular}
            \caption{Seed currents ($I_{\rm seed}$) and final runaway
              currents ($I_{\rm RE}$) for combined argon and deuterium
              injection.  $I_{\rm seed}$:
              Comparison between results using \eqref{pstarMS} and
              \eqref{pstar} with constant temperature.   $I_{\rm RE}$:
              Comparison between results using avalanche growth rate
              expressions from equations \eqref{Hava} and
              \eqref{MSava} and different assumptions for temperature
              evolution (constant, time-dependent \eqref{eq:ebalance}
              or equilibrium \eqref{eq:rad_equilibrium}). $I_{\rm RE}$
              calculated using the assumptions used in
              \citep{MartinSolis2017} (avalanche growth rate from
              \eqref{MSava}, $p_\star$ from \eqref{pstarMS} and
              constant temperature), agrees with the runaway currents
              obtained there.  Seed runaways are assumed to originate
              only from tritium decay in all cases, and $p_\star$ from
              \eqref{pstar} is used in the last three columns. }
            \label{comptab}
          \end{table}

The most  interesting aspect of the cooling in connection with major
quantities of deuterium, is the partial recombination of the deuterium due to the low temperatures. In this case, $n_\mathrm{e}^\mathrm{tot}/n_\mathrm{e}$ increases, and the screened avalanche model predicts a significantly
higher avalanche growth rate for a given electric field than in the
fully ionized case. 
The reduced ionization degree of deuterium also impacts the conductivity, as discussed in Appendix~\ref{appB}: At $1\,\rm eV$ the conductivity is reduced
by 30\% compared to the Spitzer conductivity. However, simulations with this
effect included show only a slightly higher runaway current, since the decrease in conductivity is counterbalanced by an increase in the
induced electric field. This balance is further adjusted as
the temperature also slightly increases due to a more efficient Ohmic heating for a given Ohmic current density.

So far all density profiles have been assumed to be flat. To test the sensitivity of the final runaway current to a radial variation of the density, we performed a series of simulations (not shown here) with density profiles varying from hollow to peaked.
Through the four cases considered in the paper, we have divided the parameter space into four characteristic regions in terms of the runaway dynamics. We find that, as long as the radial density variation is not so strong that different parts of the plasma would correspond to different regions in this taxonomy, both the total runaway current and the current quench time are fairly insensitive to the radial density variation. While the radial density profile does affect the runaway current profile, as long as the total number of injected atoms are held constant, the \emph{total} runaway current remains largely unaffected. A similar statement can be made regarding current quench time: the current decay rate may increase in some region and decrease in another, but total current quench time is only weakly affected. 
If, however, the radial density variation would become so strong that different parts of the plasma would correspond to different regions in our classification, effects characteristic to those regions could come into play simultaneously.

\begin{figure}
  \centering
  \includegraphics[width=0.9\textwidth]{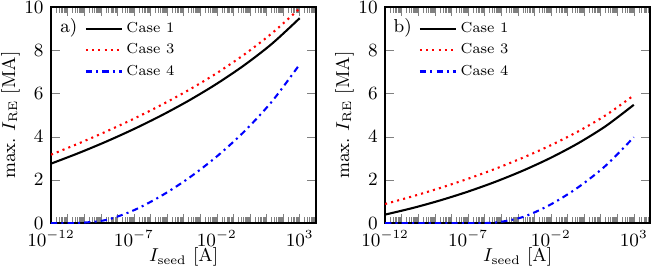}
    \caption{Maximum runaway current as function of seed current for
      Case 1, 3 and 4, assuming a flat seed runaway profile. Initial
      plasma current is (a) 15 MA and (b) 10 MA.}
    \label{fig:seed}
\end{figure}

To assess the importance of a remnant hot-tail seed, we calculate the
maximum runaway current as a function of seed current, assuming a flat
seed profile for simplicity.  Figure \ref{fig:seed} shows the maximum
runaway current as a function of seed current, for three of the cases
we considered (Case 1, 3 and 4). The final runaway current is
approximately logarithmically sensitive to the seed. The reason for
this weak dependence is that when the runaway current becomes
comparable to the Ohmic current, the electric field is reduced, which
reduces the avalanche growth rate.  In the 15 MA case, shown in
figure~\ref{fig:seed}a, as long as the seed current is above $1\;\mu
\rm A$ (corresponding to a seed runaway electron density of less than
2000 $\rm m^{-3}$) it results in more than 1 MA final runaway current
even in the most optimistic Case 4.  For a seed current of $1\;\mu \rm
A$, Cases 1 and 3 lead to 4.7 MA and 5.2 MA, respectively.

In the absence of tritium decay and Compton sources, hot-tail
generation is the only source that can give a significant runaway in
the initial, non-nuclear operational phase of ITER.  Simulations with
only Dreicer source for a 10 MA ITER scenario show that, apart from
rare cases when a filamentation of the temperature profile occurs
where a substantial Dreicer seed can be generated, the seed current is
extremely small (on the order of $10^{-10} \;\rm A$). As a result, the
total runaway current is much less than $1$ MA except for deuterium
densities high enough to cause a substantial recombination.  However,
as figure~\ref{fig:seed}b shows, if there is a remnant hot-tail seed
of the order of 0.5 A, it will result in large runaway currents also
in the 10 MA scenario (3.7 MA for Case 1, 4.3 MA for Case 3 and 1.7 MA
for Case 4). Note, that hot-tail electrons are produced under a short
period during the thermal quench, and if they are eliminated, they
will not be reconstituted. In contrast, the tritium-decay and
Compton-scattering seeds that are important in the DT phase are
produced over a longer period of time and radial losses that occur
only during the thermal quench are not sufficient for their
deconfinement.

In this paper, the plasma was assumed to be transparent to radiative
losses. However, plasmas at low temperature and high density (such as
Case 3) are partly opaque to the Lyman lines of hydrogen isotopes.
Not only is the energy balance changed due to the radiation being
trapped in the plasma, but also the ionization and recombination rates
\citep{Pshenov_2019}.  The importance of opacity effects for runaway
generation in the presence of low-Z impurities such as beryllium and
carbon has also been pointed out by \cite{Lukash_2007}. A proper
treatment of the problem requires solution of radiation transport
equations, which is outside the scope of the present paper. However,
we can estimate an upper bound for the effect of opacity by
considering the extreme case, when all the deuterium radiation is
trapped in the plasma. The temperature is then determined from the
energy balance equation \eqref{eq:ebalance}, in which the deuterium
line radiation is removed, along with the radiation corresponding to
the ionization ($P_{\rm ion}$). The latter term is important in Case
3, as shown in figure \ref{Ebalance_illustration} (note the difference
between solid and dash-dotted lines close to the left radiation
peak). Simulation results show that if only the deuterium line
radiation is trapped (i.e.~removed from equation \eqref{eq:ebalance}),
the final runaway current is reduced from $7.3 \,\rm MA$ (in the case
of a fully transparent plasma) to $6\,\rm MA$. If we also remove all
the ionization radiation in equation \eqref{eq:ebalance}, the final
runaway current is reduced to $2.73\,\rm MA$.

If we furthermore take into account the effect of opacity on the
charge states, i.e.~we use the ionization rates from the AMJUEL database\footnote{http://www.eirene.de} for the case when all
Lyman radiation is blocked in equation \eqref{eq:tdre}, we find that
recombination takes place at lower temperatures compared to the fully
transparent case. The deuterium radiation losses are then lower and
the corresponding radiation peak is moved to lower temperatures. If
line radiation losses from neutral deuterium are removed from equation \eqref{eq:ebalance}, the final runaway current becomes  $2.78\,\rm MA$, which is only slightly larger than the  $2.73\,\rm MA$ obtained when both deuterium line radiation and ionization losses are removed.

In conclusion, our estimates suggest that opacity effects can lead to significantly lower (but still substantial) final runaway currents in the case of massive deuterium injection, and should therefore be subject to further investigation.
\section{Conclusions}

Simulations of plasma shutdown scenarios indicate that impurity
injection can lead to high runaway currents in ITER. The dependence of
the avalanche generation on the density of partially ionized species
makes the runaway dynamics sensitive to the evolution of the
temperature and the distribution of ionization states. Injection of
large quantities of impurity and deuterium  results also  in a
large runaway current due to the increased avalanche generation when
deuterium recombines. On the other hand, a scenario with reduced
runaway generation was identified numerically (Case 4), for moderate
amounts of impurity and deuterium injection.

Potentially important effects that are not included in this study are  hot-tail generation and loss processes
due to magnetic perturbations, kinetic or MHD instabilities.  The
amount of runaway current deposited on plasma facing components will
depend on how large a fraction of the runaway current can be
dissipated before eventually losing the runaway beam to the wall,
which, in turn, depends on the effect of magnetic perturbations and on
how long the runaway beam can be kept stable. Thus, the effect of
transport phenomena, including the transport of neutral particles and radiation, are
outstanding issues, requiring further investigation.

\appendix

\section{Critical momentum for runaway acceleration}
\label{appA}
The critical momentum for runaway generation is sometimes determined
by setting the force balance to zero. However, this requires
either that pitch-angle scattering is negligible, or that the pitch-angle distribution
can be calculated analytically such as in \citep{AleynikovPRL}, which
is valid for electric fields near the threshold. In the case of strong electric fields and 
high plasma charge, the calculation of the avalanche growth rate by \cite{RosenbluthPutvinski1997}
can be extended to provide a threshold momentum for runaway acceleration that is valid
for an arbitrary source function,
following the same method as used in \citep{Hesslow_2019}. In this approach, which we outline below, a solution of
the kinetic equation provides an energy-dependent runaway probability, which sharply
transitions from zero to unity near a momentum $p_\star$ that we define as the critical momentum.

The kinetic equation  in the superthermal limit can be written as
\begin{align}
  \hspace{-4mm}\frac{\partial f}{\partial t} + eE_\parallel\left(\xi\frac{\partial f}{\partial p} + \frac{1-\xi^2}{p}\frac{\partial f}{\partial \xi}\right) = \frac{1}{p^2}\frac{\partial}{\partial p}\Big(p^3\nu\sub{s} f\Big) + \frac{\nu\sub{D}}{2}\frac{\partial}{\partial \xi}\left[(1-\xi^2)\frac{\partial f}{\partial \xi}\right] + S(p,\,\xi),
  \label{ke}
\end{align}
where $\xi=p_\|/p$, with the momentum component parallel to the magnetic field $p_\|$, and $S$ denotes an arbitrary particle source. We solve equation \eqref{ke}
 perturbatively in the limit of strong pitch angle scattering and strong electric fields by ordering $\nu_{\rm D} \sim \delta^0$, $eE \sim \delta$, $\nu_{\rm s} \sim S \sim \delta^2$ and $\partial/\partial t \sim \delta^3$. Then, writing $f=f_0+\delta f_1+\delta^2 f_2 + ...$, we obtain the system of equations
\begin{align}
\frac{\partial}{\partial \xi}\left[(1-\xi^2)\frac{\partial f_0}{\partial \xi}\right] &= 0, \\
eE_\parallel \left(\xi\frac{\partial f_0}{\partial p} + \frac{1-\xi^2}{p}\frac{\partial f_0}{\partial \xi}\right) &=  \frac{\nu\sub{D}}{2}\frac{\partial}{\partial \xi}\left[(1-\xi^2)\frac{\partial f_1}{\partial \xi}\right] \\
\hspace{-7mm}  eE_\parallel \left(\xi\frac{\partial f_1}{\partial p} + \frac{1-\xi^2}{p}\frac{\partial f_1}{\partial \xi}\right) &= \frac{1}{p^2}\frac{\partial}{\partial p}\Big(p^3 \nu\sub{s}f_0\Big) + \frac{\nu\sub{D}}{2}\frac{\partial}{\partial \xi}\left[(1-\xi^2)\frac{\partial f_2}{\partial \xi}\right] + S(p,\,\xi).
\end{align}
The first equation yields the general solution
\begin{align}
f_0 = f_0(t,p),
\end{align}
i.e.~the leading-order distribution is isotropic, upon which the second equation takes the form
\begin{align}
\frac{2e E_\parallel}{\nu_{\rm D} }\xi\frac{\partial f_0}{\partial p } &=\frac{\partial}{\partial \xi}\left[(1-\xi^2)\frac{\partial f_1}{\partial \xi}\right], \nonumber \\
\Rightarrow f_1 &= -\xi \frac{eE_\parallel}{\nu_{\rm D} } \frac{\partial f_0}{\partial p},
\end{align}
Integrating the third equation over $\xi$ from -1 to 1 and inserting this solution for $f_1$ yields
\begin{align}
- \frac{1}{p^2}\frac{\partial}{\partial p} \left[ p^2\left( \frac{e^2 E_\parallel^2}{3\nu_{\rm D} }\frac{\partial f_0}{\partial p} + p\nu_{\rm s}  f_0 \right)\right] = \frac{1}{2}\int_{-1}^1 S\,\rd \xi.
\label{eq:steady state}
\end{align}
from which we can derive the critical momentum, $p_\star$, as follows.
The runaway generation rate $\partial n\sub{RE}/\partial t$ can be defined as the particle flux to infinity, which, if we compare (\ref{eq:steady state}) with the initial kinetic equation, can be written
\begin{align}
\frac{\partial n\sub{RE}}{\partial t} = -4\pi p^2\left(\frac{e^2 E_\parallel^2}{3\nu_{\rm s}}\frac{\partial f_0}{\partial p} + p\nu_{\rm s} f_0\right)_{p=\infty}.
\end{align}
Therefore, if we multiply equation (\ref{eq:steady state}) by $p^2$ and integrate over all momenta, we obtain
\begin{align}
p^2\frac{e^2E_\parallel^2}{3\nu_{\rm D}}\frac{\partial f_0}{\partial p} + p^3\nu_{\rm s} f_0 = -\frac{1}{4\pi}\frac{\partial n\sub{RE}}{\partial t} + \frac{1}{2} \int_p^\infty \rd p \,p^2 \int_{-1}^1 \rd \xi \,S(p,\,\xi).
\end{align}
This first-order linear ordinary differential equation can be solved by introducing an integrating factor $G$ 
\begin{align}
G(p) &= -\int_p^\infty \frac{3p\nu_{\rm s} \nu_{\rm D}}{e^2 E_\parallel^2} \,\rd p,
\end{align}
upon which the equation takes the form
\begin{align}
\frac{\partial}{\partial p}\left( e^{G(p)}f_0(p) \right) = \frac{3\nu_{\rm D} e^{G(p)}}{p^2 e^2 E_\parallel^2}\left(- \frac{1}{4\pi}\frac{\partial n\sub{RE}}{\partial t} + \frac{1}{2} \int_p^\infty \rd p  \,p^2\int_{-1}^1 \rd \xi \,S(p,\,\xi)\right).
\end{align}
Since $p \nu_{\rm s}\nu_{\rm D} \propto 1/p^5$ for small momenta, $G \propto -1/p^4$ for $p\to 0$. Then, if we integrate this equation over $p$ from $0$ to $\infty$ and assume that $f_0$ is well-behaved (i.e.~is finite at the origin and vanishes at infinity),  we obtain
\begin{align}
\frac{\partial n\sub{RE}}{\partial t} \int_0^\infty \frac{3\nu_{\rm D} e^G}{p^2 e^2E_\parallel^2} \rd p &= \int_0^\infty \rd p \,\frac{3\nu_{\rm D} e^G}{p^2 e^2E_\parallel^2} 2\pi \int_p^\infty \rd p' \,p'^2 \int_{-1}^1\rd \xi \, S(p',\,\xi) \nonumber \\
&= 2\pi\int_0^\infty \rd p \,p^2\int_{-1}^1 \rd \xi \,S(p,\,\xi) \int_0^p\rd p' \,\frac{3\nu_{\rm D}(p') e^{G(p')}}{p'^2 e^2E_\parallel^2},
\end{align}
where in the second line we exchanged integration orders of $p'$ and $p$. From this, it follows that the runaway generation rate due to an arbitrary source function takes the form
\begin{align}
\frac{\partial n\sub{RE}}{\partial t} &= \int S(\boldsymbol{p}) h(p) \,\rd\boldsymbol{p}, 
\label{eq:final growth} \\
h(p) &= \frac{\ddint_0^p\rd p' \,\dfrac{3\nu_{\rm D}(p') e^{G(p')}}{p'^2 e^2E_\parallel^2}}{ \ddint_0^\infty \rd p' \,\dfrac{3\nu_{\rm D}(p') e^{G(p')}}{p'^2 e^2E_\parallel^2}}. \nonumber
\end{align}
Note that this equation is valid for a source term with an arbitrary pitch angle dependence and is valid for non-relativistic as well as relativistic energies.
Since $\nu_{\rm s}$ and $\nu_{\rm D}$ are monotonically decreasing functions, the function $h(p)$ will be monotonically increasing from $h(0) = 0$ to $h(\infty) = 1$, and can therefore be interpreted as the runaway probability function for an electron born at $p$ to reach arbitrarily large momenta.  

Since the function $G(p)$ varies very rapidly with $p$ --- approximately as $-1/p^4$ --- the function $h$ will make a sharp transition from essentially $0$ to $1$ in the momentum region where $G$ crosses $-1$. This can be readily confirmed for an ideal non-relativistic plasma with $\nu_{\rm s},\,\nu_{\rm D} \propto 1/p^3$, for which we can obtain the exact result $h(p) = \exp[G(p)] = \exp\{-3(1+Z)/[4(E/E_{\rm c})^2p^4]\}$.

Therefore, we may approximate $h(p)$ as a step function $h=\Theta(p-p_\star)$, where we define $p_\star$ via
\begin{align}
G(p_\star) &= -1, \nonumber \\
p_\star^2 &\approx \sqrt{\frac{3}{4}} \frac{E_\parallel/E_{\rm c}}{\sqrt{\bar\nu_{\rm s}(p_\star)\bar\nu_{\rm D}(p_\star)}} \quad \text{(non-relativistic)}.
\end{align}
The factor  $\sqrt{3/4}$ is not significant and has been neglected in this paper.  Interestingly, the expression for $p_\star$ is the same as the one appearing in the avalanche formula, that was interpreted as an effective momentum for runaway acceleration.

\section{Spitzer conductivity in weakly ionized plasmas}
\label{appB}
When the temperature in a plasma consisting mainly of deuterium
becomes as low as $1\,\rm eV$,  the ionization fraction falls to a
few percent, as shown in figure \ref{ionization}. Then the effect of
electron-neutral collisions on the conductivity has to
be taken into account.

The conductivity in a very weakly ionized plasma, where electron-neutral collisions dominate, is given by
\cite{Conductivity_neutrals} as
\begin{equation}
    \sigma=-\frac{4\pi}{3}\frac{e^2}{m_\mathrm{e}}\int_0^\infty \frac{1}{\nu_\mathrm{en}(v)}\frac{\partial f_0}{\partial v}v^3\mathrm{d}v,
    \label{eq:sigma_en}
\end{equation}
where $f_0$ denotes the Maxwell distribution of electrons, and 
$\nu_\mathrm{en}(v)=n_\mathrm{n}Q_\mathrm{en}(v)v$ is the electron-neutral
collision frequency, with the density of neutral atoms $n_{\rm n}$,   
the electron-neutral momentum exchange cross section $
    Q_\mathrm{en}=2\pi\int_0^{\pi}I(\theta,v)[1-\cos{(\theta)}]\sin{(\theta)}\mathrm{d}\theta$,
and the electron-neutral differential scattering
cross section $I(\theta, v)$.

If electron-electron collisions would be negligible in a fully ionized plasma, the conductivity would take the same form as \eqref{eq:sigma_en}, but with $\nu_\mathrm{en}(v)$ replaced by the electron-ion collision frequency, $\nu_\mathrm{ei}(v)=\Zeff (e^4 n_{\rm e} \ln \Lambda)/(4 \pi \epsilon_0^2 m_{\rm e}^2 v^3)$.
Furthermore, as shown by \citet{Spitzer}, when electron-electron collisions are accounted for, the conductivity in a fully ionized plasma with an effective ion charge $\Zeff=1$, differs from the result with electron-ion collisions only by a factor $\gamma_\mathrm{E}=0.58$.

With these considerations, \cite{Conductivity_neutrals} constructed an approximate expression for the conductivity, which reproduces the fully ionized and the very weakly ionized limits, and provides a reasonably good approximation in-between
\begin{equation}
    \sigma=-\frac{4\pi}{3}\frac{e^2}{m_\mathrm{e}}\int_0^\infty \frac{1}{\gamma_\mathrm{E}^{-1}\nu_\mathrm{ei}(v)+\nu_\mathrm{en}(v)}\frac{\partial f_0}{\partial v}v^3\mathrm{d}v.
    \label{eq:sigma_corr1}
\end{equation}

By writing $\nu_\mathrm{en}=\bar{\nu}_\mathrm{en}x\,Q(x)/\bar{Q}$ and $\nu_\mathrm{ei}=\bar{\nu}_\mathrm{ei}/x^3$, where a bar denotes a value at the thermal velocity, and $x=v/v_\mathrm{th}$, we can write \eqref{eq:sigma_corr1} as
\begin{equation}
    \sigma=\frac{\sigma_\mathrm{Sp}}{3}\int_0^\infty\frac{x^4e^{-x^2}}{\gamma_\mathrm{E}\frac{\bar{\nu}_\mathrm{en}}{\bar{\nu}_\mathrm{ei}}\frac{Q(x)}{\bar{Q}}x+\frac{1}{x^3}}\mathrm{d}x.
    \label{eq:sigma_corr2}
\end{equation}
In our calculations we use \eqref{eq:sigma_corr2} to compute the plasma conductivity, for which we obtain the momentum exchange electron-neutral collisional cross section $Q_\mathrm{en}$ for neon and deuterium from \cite{ITIKAWA197869}. Furthermore, we use the value of $\gamma_{\rm E}$ corresponding to $\Zeff=1$, since whenever the ionization degree becomes as low as to modify the conductivity significantly, the densities of the ionization states greater than 1 are low. 

\begin{figure}
  \centering
  \includegraphics[width=0.48\textwidth]{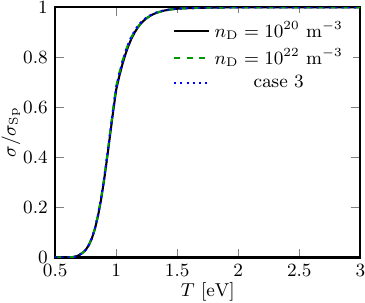}
    \caption{Electrical conductivity relative to its Spitzer value at collisional-radiative equilibrium as a function of temperature for various plasma compositions. Solid: $n_{\rm D}=10^{20} \,\rm m^{-3}$; dashed $n_{\rm D}=10^{22} \,\rm m^{-3}$; dotted $n_{\rm D}=4 \cdot 10^{21} \,\rm m^{-3}$ and $n_{\rm Ne}=8 \cdot 10^{18} \,\rm m^{-3}$ (Case 3). Note that the curves corresponding to the three cases overlap.}
    \label{fig:conductivity_correction}
\end{figure}

Figure~\ref{fig:conductivity_correction} illustrates the reduction of the conductivity compared to the Spitzer value. The ionization degree for a given temperature is calculated assuming collisional-radiative equilibrium, i.e.~by solving equation \eqref{eq:ioniz_equilibrium}. This temperature is then used to calculate the collision frequencies $\bar{\nu}_\mathrm{en}$ and $\bar{\nu}_\mathrm{ei}$, used in equation \eqref{eq:sigma_corr2} to calculate $\sigma/\sigma_\mathrm{Sp}$. While at a temperature as low as $1\,\rm eV$ the deviation from the Spitzer conductivity is only about 30\%, even if the ionization degree  is only $\approx 5\%$, if the temperature is decreased further, $\sigma/\sigma_{\rm Sp}$ drops quite rapidly.
That the three different plasma compositions considered in the figure result in essentially identical results reflects that the ionization degree is only a weak function of deuterium density, and that a comparatively small fraction of neon has a negligible effect.

\section*{Acknowledgements} \noindent The authors are grateful to S~Newton,
G~Papp, M.~Hoppe, E.~Nardon, S~Krasheninnikov and A.~Kukushkin for fruitful discussions.  This work was
supported by the Swedish Research Council (Dnr.~2018-03911), the
European Research Council (ERC-2014-CoG grant 647121), and the
EUROfusion - Theory and Advanced Simulation Coordination (E-TASC). The
work has been carried out within the framework of the EUROfusion
Consortium and has received funding from the Euratom research and
training programme 2014-2018 and 2019-2020 under grant agreement No
633053. The views and opinions expressed herein do not necessarily
reflect those of the European Commission.  \bibliographystyle{jpp}
\bibliography{references} 

\end{document}